 \documentstyle[aps,preprint,epsfig]{revtex}

 \newdimen\figwidth \figwidth=7.5cm   

 \mathchardef\dag="0279

 \def\T{{\bf T}}
 \def\e{{\bf e}}
 \def\rme{{\rm e}}
 \def\a{{\bf a}}
 \def\q{{\bf q}}
 \def\pa{{\parallel}}
 \def\pe{^\perp}
 \def\E{{\bf E}}
 \def\A{{\bf A}}
 \def\B{{\bf B}}
 \def\K{{\bf K}}
 \def\r{{\bf r}}
 \def\h{{\bf h}}
 \def\O{\relax\ifmmode{\rm O}\else\char31\fi}
 \def\H{{\bf H}}
 \def\R{{\bf R}}
 \def\Tr{\mathop{\rm Tr}\nolimits}
 \def\mmu{\setbox0=\hbox{$\mu$}\kern-.015em\copy0\kern-\wd0
	\kern.02em\copy0\kern-\wd0\kern-.015em\raise.02em\box0}

\begin{document}


 \title{Phason elasticity of a three-dimensional quasicrystal:
 transfer-matrix method}
 \author{M.~E.~J.~Newman and C.~L.~Henley}
 \address{Laboratory of Atomic and Solid-State Physics,\\
Cornell University, Ithaca.  NY 14853--2501.}
 \maketitle

 \begin{abstract}
 We introduce a new transfer matrix method for calculating the thermodynamic
properties of random-tiling models of quasicrystals in any number of
dimensions, and describe how it may be used to calculate the phason elastic
properties of these models, which are related to experimental measurables
such as phason Debye-Waller factors, and diffuse scattering wings near Bragg
peaks.  We apply our method to the canonical-cell model of the icosahedral
phase, making use of results from a previously-presented calculation in which
the possible structures for this model under specific periodic boundary
conditions were cataloged using a computational technique.  We give results
for the configurational entropy density and the two fundamental elastic
constants for a range of system sizes.  The method is general enough allow a
similar calculation to be performed for any other random tiling model. 
 \end{abstract}

 \narrowtext

 \section{Introduction} Quasicrystals are defined as structures which possess
translational order to the extent that their Fourier transforms exhibit
$\delta$-function Bragg peaks, but which have symmetries that are forbidden
in a periodic Bravais lattice.  A number of alloys have been found
experimentally which appear to be true quasicrystals in this sense, such as
i(AlCuFe)~\cite{bancel} and i(AlPdMn)~\cite{boudard,audier}, which both
exhibit resolution-limited Bragg peaks.


Two competing physical scenarios have been advanced to explain the origins of
quasiperiodic ordering.  One of them---the `ideal tiling'
scenario---postulates that the atomic structure of a well-annealed
quasicrystalline sample is perfectly quasiperiodic, like the two-dimensional
Penrose tiling~\cite{penrosegardner} or its generalizations.  In this
scenario it is hypothesized that the microscopic Hamiltonian constrains the
local arrangements of atoms in such a way as to implement something akin to
Penrose's local `matching rules', which force long-range quasiperiodicity.

The alternative to this ideal tiling approach is the `random tiling'
scenario, which also makes use of local clusters of atoms, with packing rules
similar qualitatively to Penrose's rules but insufficiently strong to force a
unique behavior at long distances.  This gives rise to an ensemble of
different ways of packing atoms into the space occupied by the sample.
Normally we assume these different configurations to be nearly degenerate in
energy.  In the actual models studied, space is assumed to be filled by a
finite set of local patterns which we represent by a set of tiles.  Such
models have a contribution to their entropy arising from the many ways in
which the tiles may be packed.  This entropy can reduce the free energy with
respect to crystalline structures (which are expected to be stable at zero
temperature), offering an alternative explanation of how the quasicrystal
state might be stabilized~\cite{widom90,henley89}.  To date, random-tiling
models have been studied in two dimensions with 8-fold~\cite{li,cockayne94b},
10-fold~\cite{widom90,binary,strandburg89} and 12-fold
symmetries~\cite{oxborrow1}, and in three dimensions with icosahedral
symmetry~\cite{tang90,shaw1,strandburg91}.

In principle it should be possible to construct random-tiling models so
that there is a one-to-one mapping of tiling configurations onto real atomic
configurations, via deterministic rules that specify how the tiles are to be
decorated with atoms.  Then the configurational entropy (and other
thermodynamic quantities) of the atomistic model will be identical to that
of the tiling model.  The decoration of random tilings to produce atomic
models of the icosahedral phases is discussed in
Refs.~\onlinecite{oxborrowthesis,mvic2,mvic3,decorate}.

This paper is principally concerned with random tiling models of
quasicrystals.  The qualitative predictions of these models and the ideal
tiling models are very similar, to the extent that they are known.  In
particular both models predict the sharp Bragg peaks observed in experiments.
Experiments aimed at resolving the differences between the two for real
quasicrystals have still not conclusively settled the question one way or the
other.  However, there are some arguments we can present in favor of the
random-tilings.  One important point is that we have some understanding of
how interatomic potentials might lead to a random-tiling structure.  The
simplest model for such a process is the 2D `binary tiling', in which atoms
of two different sizes aggregate in patterns which can be set in one-to-one
correspondence with the configurations of a tiling of `fat' and `skinny'
rhombi~\cite{binary}.  More recently, it has been shown in simulations that
several 3D models with reasonable pair potentials for simple atoms (Al, Mg,
Li, Zn) will freeze into structures which can be represented as
configurations of a random tiling model~\cite{3dpair,dzugutov}.  On the other
hand, there is no understanding of how the thermodynamics of a system
governed by a microscopic Hamiltonian might give rise to the matching rules
necessary for the ideal tiling scenario.  In fact, the most promising recent
advance in this direction has been achieved at the expense of blurring the
distinction between random tilings and ideal quasicrystal models: one assumes
an ensemble in which any of a large number of tilings is permitted, but
adopts a Hamiltonian such that the ground state, and the thermodynamic state
if the temperature is not too large, is essentially an ideal
quasicrystal~\cite{jeongsteinhardt}.



 \subsection{Phason elasticity} The long-wavelength behavior of random tiling
models is described by a sort of Landau theory, which has two kinds of
parameters: (i) the entropy density, and (ii) the `phason strain' (see
Sec.~II).  Most of the proposed experimental tests of the random-tiling
scenario hinge around the expected variation of the entropy with this phason
strain, which produces a kind of elastic term in the free energy.  We
describe this in more detail in the next section where we define the `phason
elasticity tensor', which measures the strength of this effect.  It is
predicted that the smallest eigenvalue of the phason elasticity tensor should
decrease with decreasing temperature, producing instabilities of the
quasicrystal phase with respect to other structures when the temperature
becomes low enough~\cite{henleyART,widom91,ishii}.

Also, by contrast with the matching rule models, it is predicted that the
phason elasticity for a random-tiling should have a gradient-squared form
(see Section~II~A).  This leads to diffuse `wings' around peaks in the
diffraction pattern, and to phason contributions to the Debye-Waller
reduction of the Bragg intensities, both of which can in principle be
measured experimentally to determine values for the elastic constants.

 \subsection{Random canonical-cell tilings} In this paper, we build upon
results from our previously published work~\cite{newman1}, henceforth
referred to as Paper I, to calculate a variety of properties of a random
tiling ensemble based on the `canonical cell' model of a quasicrystal.  This
model has been described in detail by Henley~\cite{henleyCCT}.  Briefly, it
consists of vertices or `nodes' which represent icosahedrally-symmetric
clusters of atoms.  Nearest-neighbor nodes are joined by `linkages' of two
types.  Type $b$ linkages run parallel to the axes of two-fold symmetry of
the reference icosahedron and are all of the same length which we denote $b$;
type $c$ linkages run parallel to the axes of three-fold symmetry, and all
have length $c\equiv b\sqrt3/2$.  The linkages can form three kinds of
polygon which in turn form the faces of four kinds of tiles or `canonical
cells' $A$, $B$, $C$, and $D$ into which the entire space is divided.  (These
cells were depicted in Fig. 1 of Paper~I and also Fig. 1 of
Ref.~\cite{henleyCCT}.)



In Paper~I we gave an algorithm for generating a so-called `stacking graph'
for a random tiling, which we applied to the particular case of the
canonical-cell tiling.  This graph contains information about all possible
structures that can be built out of a certain set of tiles with specified
periodic boundary conditions.  In this paper, we use this stacking graph as
the basis for a transfer matrix calculation of the thermodynamic properties
of the ensemble of random tilings of canonical cells, including the random
tiling entropy and the phason elasticity~\cite{newman2}.  These quantities
have been found previously for another icosahedral random tiling, that of the
two Ammann rhombohedra~\cite{strandburg91}.  The canonical-cell model,
however, is different in at least two respects:
 \begin{itemize}
 \item[(i)] It lends itself to the construction of realistic atomic models
for icosahedral quasicrystals such as i(AlCuLi)~\cite{mvic2},
i(AlMnSi)~\cite{decorate}, i(TiCrSi)~\cite{mvic3}, and potentially
i(AlPdMn).  It is possible to construct decorations of the canonical cells
which incorporate our understanding of the basic atomic motifs found in
these alloys and which contain no points of unrealistically close or loose
packing.  Conversely it has not proved possible to relate the rhombohedron
tiling to any good structure model. 
 \item[(ii)] The canonical cell model is, unfortunately, much less tractable
technically than the tiling of Ammann rhombohedra.  In particular, Monte
Carlo simulation has not been practicable because there is no known move that
rearranges tiles locally and satisfies ergodicity.  A possible chain-like or
cluster move has been suggested by Oxborrow~\cite{oxborrowthesis}, but this
move is still not completely understood.  (It has been understood in a
two-dimensional toy model, the square-triangle random
tiling~\cite{oxborrow1}, which shares some features with the canonical-cell
tiling.)  Since we suspect that this kind of intractability is more generic
than the behavior of rhombus and rhombohedron
tilings~\cite{oxborrow1,cockayne94a,mvic4}, it behooves us to develop
appropriate methods for studying less agreeable tilings such as the
canonical-cell tiling, even if some other tiling ultimately proves more
relevant to real quasicrystalline phases.
 \end{itemize}

 \subsection{Outline of the paper} The paper is organized as follows.  In
Section~II we develop the elastic theory of random tiling ensembles,
including the special cases that apply to the ensembles we will be studying
with our transfer matrix method.  In Section~III we define our transfer
matrix and explain how it is used to calculate thermodynamic properties of
the ensemble as a function of phason strain.  In Section~IV we describe the
calculations we have performed using the canonical-cell tiling, and give our
results for the configurational entropy and phason elasticity of the tiling
for a variety of system sizes.  In Section~V we present our conclusions.

 \section{Phason elasticity in random tilings} In studying icosahedral
tilings, it is convenient to use a basis of unit vectors $\e^\pa_\alpha$
pointing from the center to vertices of the reference icosahedron,
i.e.,~along the fivefold symmetry axes.  It turns out that every type $b$
linkage can be written as the sum of four of these basis vectors and every
type $c$ linkage can be written as the sum of three of them.  Thus, having
arbitrarily assigned $\r=0$ at one node, every other node in a
canonical-cell tiling may be represented in the form
 \begin{equation}
 \r = \sum_{\alpha=1}^6 n_\alpha \e^\pa_\alpha,
 \label{defsr}
 \end{equation}
 where $n_\alpha$ are integers.  The $\lbrace\e^\pa_\alpha \rbrace$ are
independent over integers so this representation is unique.

 We can also define a `phason' coordinate for each node
 \begin{equation}
 \r^\perp = \sum_{\alpha=1}^6 n_\alpha \e\pe_\alpha.
 \label{defsh}
 \end{equation}
 The vector $\r^\perp$ lives in a space called `phason' space or `perp'
space.  As discussed elsewhere~\cite{henleyART}, with a proper choice of
basis vectors $\lbrace\e\pe_\alpha\rbrace$ the perp-space coordinate is
well-defined.  We can view Equations~(\ref{defsr}) and~(\ref{defsh})
together as defining a `lifting' of the nodes to points on a six-dimensional
hypercubic lattice; thus an arbitrary tiling configuration may be viewed as
the projection of a three-dimensional surface embedded in
six-space~\cite{elser85b}.  Following the conventions of
Jari\'c~\cite{jaricnelson}, we write
 \begin{eqnarray}
 {\e^\pa}_1 &=& \eta (\tau,0,1),\qquad
 {\e^\pa}_2  =  \eta (\tau,0,-1),\qquad
 {\e^\pa}_3  =  \eta (1,\tau,0),\nonumber\\
 \e^\pa_4   &=& \eta (0,1,\tau),\qquad
 \e^\pa_5    =  \eta (0,-1,\tau),\qquad
 \e^\pa_6    =  \eta (1,-\tau,0),
 \label{parallel}
 \end{eqnarray}
 and
 \begin{eqnarray}
 {\e\pe}_1 &=& \eta (1,0,-\tau),\qquad
 {\e\pe}_2  =  \eta (1,0,\tau),\qquad
 {\e\pe}_3  =  \eta (-\tau,1,0),\nonumber\\
 {\e\pe}_4 &=& \eta (0,-\tau,1),\qquad
 {\e\pe}_5  =  \eta (0, \tau,1),\qquad
 {\e\pe}_6  =  \eta (-\tau,-1,0),
 \label{perp}
 \end{eqnarray}
 where $\tau \equiv {1 \over 2} (1 + {\sqrt{5}})$ and $\eta\equiv (1 +
\tau^2)^{-1/2}$~\cite{choice}.  With these definitions, the length of the
$b$-type linkage is $b=2\sqrt{1+2/\sqrt5}$.  We also define a coarse-grained
phason coordinate
 \begin{equation}
 \h(\r) =  \langle \r\pe \rangle_\r
 \label{coarse}
 \end{equation}
where $\langle \ldots \rangle_\r$ is an average over a neighborhood of $\r$.
The phason strain tensor is then the gradient of $\h$ in real space:
 \begin{equation}
 \E(\r) \equiv \nabla_{\r} \h (\r). 
 \label{defse}
 \end{equation}

 \subsection{Form of the elasticity tensor} Consider an ensemble of tilings,
possibly weighted with Boltzmann probabilities derived from some Hamiltonian
${\cal H}$.  We hypothesize that the Hamiltonian is sufficiently small, or
the temperature sufficiently high, that difference in energy between
configurations is typically much less than $kT$, so that the free energy of
the ensemble will be dominated by the entropy.  We further make the
assumption that the relative weighting of long-wavelength phason fluctuations
(after coarse-graining) will be determined by a free energy of the
form~\cite{elser87,henley87}
 \begin{equation}
 {\cal F} = \int_V d^3\r f(\E(\r))
 \end{equation}
 where
 \begin{equation}
 f(\E)= f_0 + {1\over 2} \E\cdot {\bf K} \cdot \E + \O(\E^3).
 \label{freal}
 \end{equation}
 ${\bf K}$ is the elasticity tensor (really a 4-tensor) and $V$ is the system
volume.  In the case where the Hamiltonian ${\cal H}$ is zero for all states
in the ensemble, $f_0$ is equal to minus the entropy per unit volume
$S_V$~\cite{tisone}.  It is important to be aware that this form for the free
energy is just a hypothesis.  We cannot prove that $f(\E)$ must have an
analytic minimum at $\E=0$, though it can be shown that, if the linkages in
our tiling are uniformly distributed over all orientations equivalent by
icosahedral symmetry, then $\r^\perp$ is approximately constant.  In other
words, the hypersurface formed by the tiling in six-dimensional space is
approximately flat, having only small phason fluctuations at non-zero
wavevectors.  Thus the phason strain can be thought of as parametrizing the
deviation from icosahedral symmetry, making it natural for $f$ to have a
minimum at $\E=0$ if the free energy is dominated by the entropy of the
ensemble.  It would in theory be possible for $f$ to have a non-analytic
minimum at $\E=0$.  However, in all models previously studied the analytic
form~(\ref{freal}) as been borne out both by analytic calculations and by
simulations~\cite{tang90,shaw1,jeong}.  Furthermore, a random tiling has
long-range order (i.e.,~Bragg peaks exist) if and only if $\h(\r)$ has finite
variance; this can be shown to follow from the gradient-squared
form~(\ref{freal}).

It is common to deal with~(\ref{freal}) by assuming that the net phason
strain is zero and then re-expressing~(\ref{freal}) in terms of Fourier
components $\h(\q)$ (this generalizes trivially to the case of any fixed
background phason strain---see Section~II~B below).  The result is a sum over
wavevectors $\q$ of terms each of which is a quadratic form in the three
components of $\h(\q)$.  The coefficients are linear in $\{ K_i \}$ and
quadratic in the components of $\q$.  This form is simpler
than~(\ref{freal}), since we write a quadratic form in three instead of nine
quantities (and, what is more, the different $\q$ vectors decouple), and is
appropriate for analyzing fluctuations from Monte Carlo simulations.
However, the ensemble we will be working with in the present study permits
undetermined phason strains with respect to only the (real space) $z$
direction, and we measure only the elasticities associated with that
direction.  In this case the Fourier analysis is not useful, and we must
struggle with the full form of the elastic free energy.

  

The elastic free energy for icosahedral symmetry can be
written~\cite{jaricnelson,lubensky88,jaric87}
 \begin{eqnarray}
 f(\E) &=& {1\over2} K_1\sum_{ij} E_{ij} E_{ij} + {1\over 2} K_2 \lbrace
 (E_{11}+E_{22}+E_{33})^2 - {4\over 3} \sum _{ij} E_{ij}E_{ij} +\nonumber\\
 & & (\tau E_{12} + \tau^{-1} E_{21})^2 + \mbox{cyclic permutations}\rbrace
 + \O(\E^3).
 \label{fico}
 \end{eqnarray}


If the icosahedral phase is to be stable against decomposition into other,
possibly periodic phases, the quadratic stationary point in the free energy
must be a minimum, rather than a saddle point or maximum.  This requires that
$K_1$ be positive definite, and $K_2$ may have either sign but must lie
within the range
 \begin{equation}
 -\frac35 < K_2/K_1 < \frac34.
 \label{stability}
 \end{equation}
 The values of $K_1$ and $K_2$ are one of the main points of contact
between quasicrystal experiments and random-tiling theory.  Knowing merely
the ratio $K_2/K_1$ between the two is enough to determine the form of the
diffuse scattering from the random tiling.

 \subsection{Background phason strain} Consider now the case of a tiling
with periodic boundary conditions in all three real space directions.  The
periodicity of the structure is represented by a set of vectors
$\lbrace\a_i\rbrace$, which are the displacement in real space from a node
to the corresponding node in another periodic repetition of the structure.
Since the vectors $\a_i$ connect nodes, they can be written in the
form~(\ref{defsr}), each with a corresponding perp displacement
${\a\pe}_i$.  The boundary conditions then constrain the system to have a
global average phason strain $\B$, defined by the linear system of equations
 \begin{equation}
 \B\cdot\a_i = {\a\pe}_i.
 \label{background}
 \end{equation}
 It will still be possible to consider fluctuations in the phason strain at
finite wavevectors around this state, but when we consider the free energy
associated with the phason elasticity, we must allow for the constraint on
the long-wavelength phason strain by introducing Lagrange multipliers.  We
will not follow through the complete analysis here, but merely quote the
important results, which are
 \begin{itemize}
 \item[(i)] The elastic free energy density $f$ is shifted by a constant term
$f_0 = {1\over 2} \B\cdot\K\cdot\B$.
 \item[(ii)] The quadratic term becomes ${1\over 2} (\E-\B) \cdot {\bf K}
\cdot (\E-\B)$.
 \item[(iii)] Most significantly, cubic and higher anharmonic terms now give
contributions to the new quadratic term.  Thus the elasticity tensor in the
presence of the background phason strain loses its icosahedral symmetry and
retains only the symmetry of the Bravais lattice defined by the periodic
boundary conditions.  The deviations, however, will be at most of $\O(\B)$
(from the cubic terms).  Such terms in the elasticity were first measured
by Oxborrow~\cite{oxborrow1}.
 \end{itemize}

 \subsection{Elasticity theory for a tower} In most of this paper, we
consider a `tower' in which periodic boundary conditions are applied in only
the $x$ and $y$ directions; in the $z$ direction the tower can be
arbitrarily large.  The towers we will be considering have a square base
and, of the icosahedral symmetry operations, only $x$, $y$, and $z$
reflections will be symmetries of the ensemble of possible configurations of
cells in the tower; in particular, notice that the $x$ and $y$ axes will be
inequivalent, even though the dimensions of the base in the $x$ and $y$
directions are equal, since the icosahedral point group has no 4-fold axis
of symmetry.

For such a tower, only gradients with respect to the real-space $z$
coordinate remain free.  Gradients with respect to $x$ and $y$ are fixed by
the periodic boundary conditions, giving rise to a background phason strain
of the type discussed above.  We define $\B$ for a tower just as before
except that the components $\B_{\alpha 3}$ are zero.  Then
 \begin{equation}
 B_{11} = B_{22} = B
 \end{equation}
 with 
 \begin{equation}
 B=\tau^{-(2n+3)},
 \label{bps}
 \end{equation}
 where $n$ is the order of the approximant, with $n=1$ for the $\frac21$
structures, $n=2$ for the $\frac32$, and so forth.  All other components of
$\B$ are zero.  For the phason strain tensor $\E$, the boundary conditions
give us just three free components $E_{\alpha3}$ out of the nine.  These
three components transform like a vector, which we will call $\A$, with
 \begin{equation}
 A_\alpha \equiv d h_\alpha/dz \equiv E_{\alpha 3}.
 \end{equation}
 We can write the free energy density for the tower in terms of this vector
as
 \begin{equation}
 f(\A) = f_0 + \frac12\left\lbrace C_1 A_1^2 +  C_2 A_2 ^2 + C_3 (A_3 -
 A_{3}^{(0)})^2\right\rbrace.
 \end{equation}
 There are no cross-terms, because of the reflection symmetry, but the
linear term shown involving the constant $A_3^{(0)}$ is possible because the
$z$ reflection symmetry reverses both $h_3$ and $z$. 

Using Equation~(\ref{fico}), we can show that the constants in this formula
are given by
 \begin{eqnarray}
 C_1 &=& K_1 +(\tau^{2}-\frac43) K_2\nonumber\\
 C_2 &=& K_1 +(\tau^{-2}-\frac43) K_2\nonumber\\
 C_3 &=& K_1 - \frac13 K_2
 \label{defsc}
 \end{eqnarray}
 and
 \begin{equation}
 A_{3}^{(0)} = -2 B K_2/(K_1 - \frac13 K_2).
 \end{equation}


We will calculate values for $C_1$, $C_2$, $C_3$, and $A_3^{(0)}$ using a
transfer matrix method.  Since these quantities are all functions of two
elastic constants $K_1, K_2$, there is some redundancy in such a calculation,
which allows us a check on the accuracy of our methods.  Inaccuracies can be
ascribed to either (i) finite-size effects or (ii) finite-$B$ effects
(through the cubic terms in the elasticity theory, which we omitted).



 \subsection{Frequencies of objects in tiling} The relative frequency with
which various types of tiles appear in a random tiling is directly related to
the phason strain.  For example, in the case of certain extreme phason
strains, the tiling can be composed entirely of one type of tile.  In some
random tilings, the number densities of the different tile types is fixed
exactly by the phason strain.  The two-dimensional Penrose tiling of rhombi
is one such example.  In the canonical cell tiling this is not the case, but
there are still things we can say about the relationship of the number
density of tiles to the phason strain.  These topics were discussed in detail
in Ref.~\onlinecite{henleyCCT}.  Here, we will just summarize the situation.

In addition to the phason strain, we can define many other macroscopic
variables as the density (per unit volume) of occurrences of selected local
patterns in the tiling.  There are two reasons why we are interested in such
patterns:
 \begin{itemize} 
 \item[(i)] When we come to decorate our tilings with real atoms to create
possible atomic structures for icosahedral alloys, the simplest choice we can
make is to use a single decoration for each type of canonical cell.  In some
cases a more satisfactory structure can be produced using a
``context-dependent'' rule, in which the decoration of a tile depends on the
those of the tiles surrounding it.  In either case the decoration depends on
the local patterns in the tiling and the density and stoichiometry of the
resulting structure are simple functions of the densities of those patterns.
 \item[(ii)] As discussed below in Section~III~D, iteration of the
transfer-matrix always generates an ensemble which maximizes the entropy {\em
per layer}, whereas we actually want the ensemble which maximizes the entropy
{\em per unit volume}.  We will be introducing chemical potentials which
couple to the densities of local patterns in the tiling.  By varying these
chemical potentials, we can vary the relative frequency in the tiling of the
various patterns and so generate an ensemble which is closer to the desired
one~\cite{chempots}.
 \end{itemize}

The simplest and most important of the density variables are the densities of
the tiles $A$, $B$, $C$, and $D$.  In fact, these densities can be
parameterized by just one independent parameter $\zeta$, which is defined as
the volume ratio $V(D)/[V(A)+V(D)]$, where $V(X)$ is the total volume
occupied by tiles of type $X$.  (This follows because $V(B)$ and $V(C)$ are
unique functions of the phason strain---see Eqs.~(3.1) and~(3.6) of
Ref.~\onlinecite{henleyCCT}.)  This is important not only because $\zeta$
determines the stoichiometry and density of simple decoration models, but also
because it completely determines the density of nodes, i.e.,~the frequency of
the cluster motif in those models.  Furthermore, the bulky $D$ cell tends to
force its surroundings, so increases in its density tend to decrease the
entropy density and increase the volume added per dead-surface (see
Sec.~III~A below).  Thus, the chemical potential conjugate to $\zeta$ is the
most important one for approaching the correct ensemble.

We can write a more general form for the free energy $f(\E;\zeta)$ which has
an absolute maximum at $\E=0, \zeta=\zeta_0$.  (Possibly it might depend on
additional density parameters also, but for the moment we will stick to this
simplest case.) Then the expansion of $f$ around this maximum will contain
terms such as ${1\over 2} K_\zeta (\zeta-\zeta_0)^2$, where $K_\zeta$ is a
new elastic constant.  (Cross terms between $\bf E$ and $\zeta$ are forbidden
by symmetry.)  In the present calculations, this extra elastic constant is
not physically relevant, for reasons which we discuss in Section~III~D below,
so we define the entropy as the maximum $S(\E)=\max_\zeta[S(\E,\zeta)]$,
making it a function only of the phason strain.

 \section{Transfer matrix approach} In the majority of previous calculations
on three-dimensional random tilings, workers have made use of Monte Carlo
methods to study the elastic properties.  In two dimensions on the other
hand, the transfer matrix method has probably been as important as Monte
Carlo simulation~\cite{li,widometal89,shaw2,widom93}.  General views of the
transfer matrix method as applied to two-dimensional quasicrystals may be
found in Ref.~\onlinecite{henleyART}, Sec.~8.5.1, and in
Ref.~\cite{henley87}.  Transfer matrices provide a potentially exact way to
calculate phason elasticity, and are more convenient for calculating
entropies, which can be extracted from a Monte Carlo simulation only after
integrating an entropy differential from zero to infinite
temperature~\cite{oxborrow1,strandburg91,tang89}.  In the case of the
twelve-fold `square-triangle' tiling~\cite{oxborrow1}, a transfer-matrix
formulation has made possible an exact solution with analytic expressions for
the entropy density and the elastic constants~\cite{widom93,kalugin94}.

In the case of the canonical cell tiling, the implementation of a Monte
Carlo simulation has, as noted above, been blocked by the lack of an update
move.  So instead, we have applied the transfer matrix method, using a new
technique, which we now describe.

As described in Section~II~C, we consider an ensemble of all configurations
filling a `tower' with a given, finite base and having periodic boundary
conditions in the transverse ($x$ and $y$) directions, but unbounded in
the $z$ direction.  This reduces our three-dimensional problem to one which
is essentially one-dimensional, and therefore amenable to a transfer matrix
treatment.  We specialize to the `maximally random' ensemble of
tilings~\cite{henleyART}, meaning that each tiling configuration has equal
weight (the Hamiltonian ${\cal H}$ is zero). 

In Paper~I we demonstrated how our towers can be decomposed into layers, and
how the set of all possible towers may be codified in a `stacking graph',
which represents all the different ways of stacking these layers one on top
of another.  In this section we review this development and then discuss how
the stacking graph may be turned into a transfer matrix whose eigenvalues are
related to the free energy of the ensemble of tilings.  We will also have to
deal with a few technicalities associated with the facts that (i) the layers
do not have equal volume and (ii) we must be able to compute the free energy
for any accessible phason strain.

\subsection{Dead-surfaces and the stacking graph} The basis of our
transfer-matrix approach is a rule for representing any given tower of cells
by a sequence of layers of cells stacked in the $z$ direction, with a
one-to-one correspondence between possible tilings and sequences of layers.
Given that we can formulate a rule for dividing our tower of cells into
these layers (possible strategies for doing this are discussed below) we can
then construct a `stacking graph' for the complete ensemble of tilings.  The
stacking graph consists of vertices joined by arrows (see Fig.~4 of
Paper~I, for example) where each vertex represents one kind of layer, and an
arrow from one vertex to another represents a possible way in which the
first kind of layer can be followed by the second.  Any possible structure
can then be described by a particular path through the stacking graph.

There are two different approaches to dividing a system into layers.  In the
first approach, which has been used in all previous transfer matrix studies
of random tilings, one slices the structure up into {\em slabs\/} of cells,
which span the entire cross-section of the tower from side to side.  The
average number of cells in a slab thus scales with the cross-sectional
area.  In two dimensions, the layers can often be mapped onto rows of sites
on some lattice.  By interpreting the stacking direction as a `time' axis,
conditions on the allowed transitions from one layer to the next can then
be interpreted as local rules for the hopping of a conserved set of
particles moving in one dimension.  Making use of this idea for the case of
the square-triangle tiling, a Bethe ansatz solution has been
found~\cite{widom93}, yielding exact analytic values for the entropy and
elastic constants~\cite{kalugin94}. 


The decomposition of three-dimensional tilings into slabs is much less
promising since each layer is itself a 2D random tiling spanning the
cross-section of the tower.  At every step in the construction of the
stacking graph, we would need to completely enumerate all the possibilities
for the next layer, and the number of such possibilities would grow
exponentially as a function of the cross-sectional area, making the stacking
graph extremely convoluted.  So in Paper~I we introduced a different
approach, in which towers of canonical cells are divided into layers
separated from one another by `dead surfaces'~\cite{deadsurf}.  Using this
method, we generated the stacking graph and applied it to the production of
an exhaustive list of periodic structures on the square `$\frac53$'
base~\cite{newman1}.

The notion of a `dead surface' is linked with that of `forcing'.  Imagine
that we build a tower of cells, starting from some initial base, by adding
nodes one at a time to the open top surface.  Often, the top surface will
have certain crevices which {\em force\/} the addition of a node at a
particular place.  The reason for this is that there are a number of linkages
leaving each vertex in the surface, and there are only a finite number of
different ways in which linkages are allowed to meet at a vertex.  It is
possible therefore, and indeed quite common, for the linkages which are
already in place to restrict the number of possible choices for the ways in
which a vertex can be completed.  Thus the addition of a linkage or
linkages around that vertex may be `forced', and with these forced linkages
come new vertices that sit at their other ends.  These are our forced nodes.

\begin{figure}
 \hbox to \textwidth{\psfig{figure=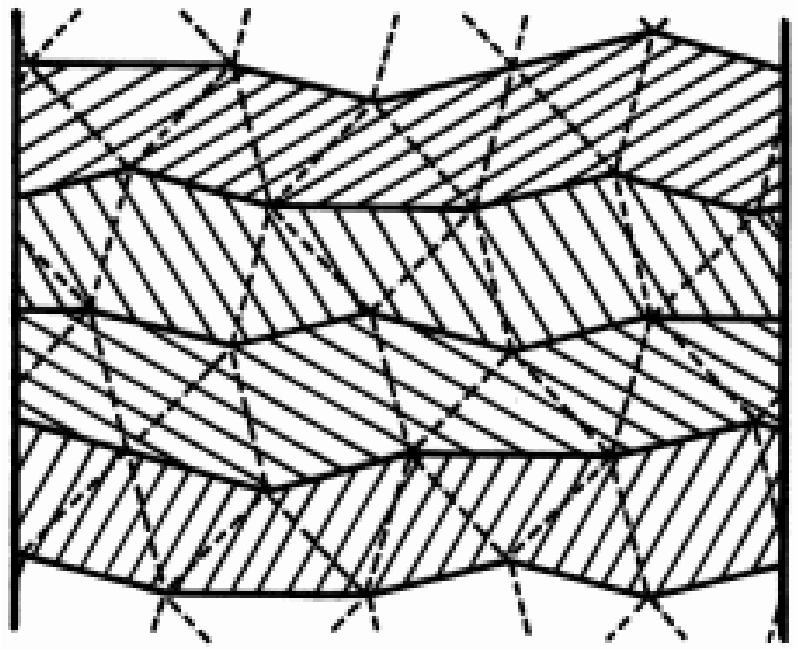,width=\figwidth}\hfil
                     \psfig{figure=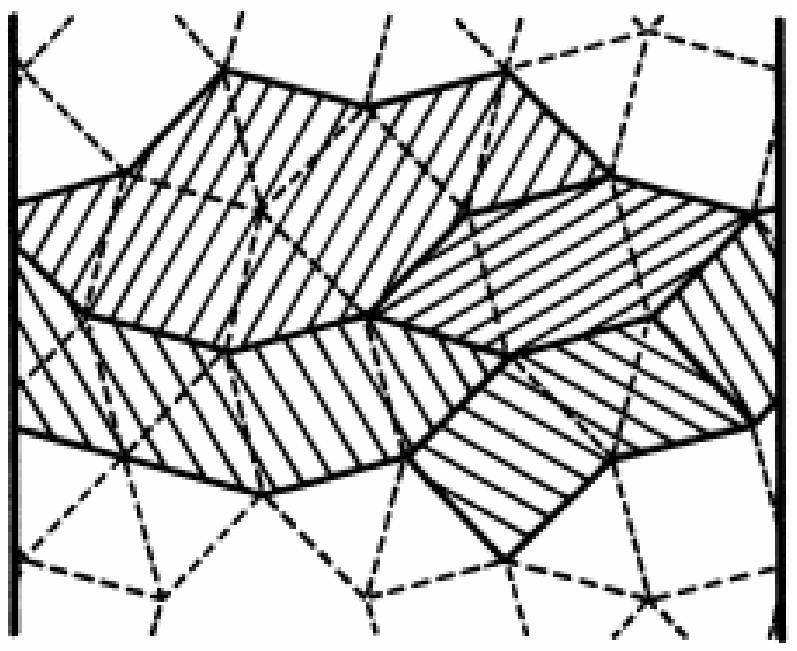,width=\figwidth}}
 \caption{Dividing a strip of square-triangle tiling into layers.  (a) `Slab'
approach (as used by~\protect\cite{oxborrow1}). (b) `Dead-surfaces' approach
(as used in the present paper for the 3D canonical-cell tiling.}
 \label{layers}
 \end{figure}

A surface which has {\em no\/} forced nodes around any of its vertices is
called a `dead surface'.  We can produce a dead surface by taking a tower of
canonical cells and adding all forced nodes to the top until no more
additions are forced.  The only choices we have to make in the construction
of the tower are the ones we make about which node to add next when we are at
a dead surface.  Thus, a list of the successive dead surfaces and the choices
we made when we got to them are sufficient to specify the tiling uniquely.
The typical number of cells between two successive dead surfaces is
relatively small, and approaches a small limit as the cross-sectional area
of the tower diverges, so that the work involved in finding a surface does
not increase indefinitely with the size of the system.  And, most important,
the stacking graph is sparse, in the sense that there only a small number
(usually two) of possible successors to a given dead surface (by contrast
with the slabs approach, in which there are exponentially many as the area of
the slabs becomes large)~\cite{isingtm}.  The `dead surface' and `slab'
approaches are contrasted in Fig.~\ref{layers}, using the two-dimensional
square-triangle tiling for pedagogical purposes.

\subsection{Construction of the transfer matrix} If we were simply to
generate tilings at random by taking random paths through the stacking graph
and reconstructing the towers of cells to which they correspond, we would not
generate tilings with the same weights with which they appear in our
ensemble.  To see this, consider a stacking graph in which, say, layer 1 can
lead to two different layers, numbered 2 or 3; taking a random sequence means
we would go to layer 2 or 3 with equal probability.  But in the random-tiling
ensemble, it may be that layer 2 is more likely than layer 3.  That will be
the case if layer 2 ultimately leads to a larger number of tilings afterwards
than layer 3 does.  If we specify that all tilings should appear with equal
weights (${\cal H}=0$), that is not the same as saying that all transitions
from one layer to another appear with equal weights.  It is in order to
tackle this problem that we introduce the transfer matrix.

As explained in the preceding section, a tiling of $M$ layers is uniquely
represented by a sequence of dead surfaces and the choices made to continue
growing the tiling at each surface.  We will denote this $(\sigma_m, l_m)$
with $m=1, \ldots, M$ where $\sigma_m$ is the label of the $m^{\rm th}$ dead
surface and $l_m$ is the label of the transition $\sigma_m\to\sigma_{m+1}$ in
the stacking graph.  (Normally there will only be one choice we can make at
surface $\sigma_m$ that will lead us to surface $\sigma_{m+1}$.  But
occasionally there will be more than one way to make the transition, and in
these cases we use the label $l_m$ to distinguish between them.)

We also partition the Hamiltonian into terms which each involve just
two surfaces and the tiles contained between them:
 \begin{equation}
 {\cal H} = \sum _{m=1}^M U(\sigma_m,\sigma_{m+1},l_m).
 \label{hamilayer}
 \end{equation}
Then the partition function can be written as 
 \begin{equation}
 {\cal Z} = \sum_{\lbrace\sigma_m,l_m\rbrace} \rme^{-{\cal H}} = \Tr\T^M,
 \label{defstm}
 \end{equation}
 where $\T$ is the transfer matrix. The general form for the elements
of the transfer matrix is
 \begin{equation}
 T_{\sigma\tau} \equiv \sum_l\rme^{-U(\sigma,\tau,l)}.
 \label{telements}
 \end{equation}
 Note that $T_{\sigma\tau}$ is zero if there is no connection $\sigma\to\tau$
in the stacking graph; for the particular case ${\cal H}=0$, $T_{\sigma
\tau}$ is just the number of ways to get from $\sigma$ to $\tau$.

The total weight in the partition function of all configurations of $M$
layers starting with layer $\sigma$ and ending with layer $\tau$ is then
$(\T^M)_{\sigma\tau}$.  It can be seen that, whatever boundary conditions we
choose for the first and last layers, the free energy {\em per layer\/} in
the thermodynamic limit is 
 \begin{equation}
 F_{layer} = -\ln\Lambda_{max}
 \label{defsg}
 \end{equation}
 where $\Lambda_{max}$ is the largest eigenvalue of $\T$. 

We are interested in computing the free energy for a range of phason
strains, and also for a range of values of the parameter $\zeta$ (see
Section~II~D).  This means we must consider (i)~how to measure the mean phason
strain and density of a given ensemble, and (ii)~how to generate ensembles
with differing phason strain and density.  These are dealt with in the next
two subsections. 

 \subsection{Expectations} Having found the dominant eigenstate of the
transfer matrix, we can extract the expectation of any operator ${\cal Q}$
that can be written in the form of a sum over the transitions from one dead
surface to the next:
 \begin{equation}
 {\cal Q} = \sum _{m=1}^M Q(\sigma_m,\sigma_{m+1},l_m).
 \label{operator}
 \end{equation}
 If $v^{(L)}$ and $v^{(R)}$ are the left and right eigenvectors
corresponding to $\Lambda_{max}$, then the expectation of ${\cal Q}$ is
 \begin{equation}
 \langle {\cal Q} \rangle = \sum_{\sigma,\tau,l}
 v^{(L)}_\sigma \rme^{-U(\sigma,\tau,l)} Q(\sigma,\tau,l) v^{(R)}_\tau.
 \end{equation}
Thus, for example, the expectation of the Hamiltonian, which is just the
average energy per layer of the random tiling is
 \begin{equation}
 \langle U \rangle = \sum_{\sigma,\tau,l}
 v^{(L)}_\sigma \rme^{-U(\sigma,\tau,l)} U(\sigma,\tau,l) v^{(R)}_\tau. 
 \end{equation}

A number of other useful operators besides the Hamiltonian may be written in
the form~(\ref{operator}).  In particular, having designated a representative
node on each dead surface, we can define vectors $\R(\sigma,\tau,l)$ and
$\H(\sigma,\tau,l)$ which are respectively the step in physical space and in
perp space from one dead surface to the next.  Then the total offset in
physical and perp space are given by
 \begin{equation}
 \R_{tot} = \sum_{m=1}^M \R(\sigma_m,\sigma_{m+1},l_m)
 \end{equation}
 and
 \begin{equation}
 \H_{tot} = \sum _{m=1}^M \H(\sigma_m,\sigma_{m+1},l_m).
 \end{equation}
 In the thermodynamic limit, the phason strain tensor $\E$ is then given by
 \begin{equation}
 \H_{tot}= \E\cdot\R_{tot}.
 \label{htot}
 \end{equation}
 Six of the nine components of $\E$ take fixed values which we already know
(see Section~II~C).  The remaining three we calculate from~(\ref{htot}).
   
We can also write the total number of nodes as
 \begin{equation}
 N_{tot} = \sum _{m=1}^M N(\sigma_m,\sigma_{m+1},l_m).
 \label{nodelayer}
 \end{equation}
 where $N(\sigma_m,\sigma_{m+1},l_m)$ is the number of nodes added between
surfaces $m$ and $m+1$.  Since the total volume is given by 
 \begin{equation}
 V_{tot} = [\R_{tot}]_z a^2
 \label{volume}
 \end {equation}
 (where $a$ is the length of the edge of the square base), the number
density of nodes is given by the ratio $N_{tot}/V_{tot}$.


 \subsection{Chemical potentials} We wish to investigate ensembles of towers
of canonical cells with different mean phason strains, so as to be able to
maximize the random tiling entropy as a function of phason strain.  In order
to vary the average values of the three free phason strain components in our
ensemble, we need to introduce terms into our Hamiltonian which couple to the
phason strain, as mentioned briefly in Section~II~D.  For example, if we want
a particularly large $E_{xz}$ component of phason strain, we should favor
those transitions from one surface to another which contribute a large shift
in the $x$ direction in perp space by comparison with the $z$ component of
the accompanying real space shift.  The appropriate form for the terms in the
Hamiltonian to achieve this is $\mmu\cdot\A$, where $\A$ is the vector
introduced in Section~II~C composed of the three free strain components, and
$\mmu$ is a vector whose components are chemical potentials coupling to the
phason strain.  Then the entropy per layer $S_{layer}$ is given by Legendre
transformation:
 \begin{equation}
 F_{layer} = -\ln \Lambda_{max} = -S_{layer} + \mmu\cdot\overline\A
 \end{equation}
 where $\overline\A$ is the mean phason strain.  Physically, we are only
interested in the case ${\cal H}=0$; the chemical potentials are added solely
as auxiliary fields in order to probe the variation of the entropy with phason
strain.

We can similarly introduce a chemical potential $\mu_\zeta$ coupling to the
the parameter $\zeta$, or equivalently to the packing fraction of nodes.
However, we assume that $\zeta$ is freely varying, since we do not envisage
placing any physical constraint on our ensemble that would fix its value.  So
the appropriate course of action will always be to choose the value of
$\mu_\zeta$ that maximizes the entropy density~\cite{muphi}.



For a system which did have a non-zero Hamiltonian, the equilibrium value of
the internal energy (where $T=1$) would be given by
 \begin{equation}
 U = S - \mmu\cdot\overline\A - \mu_\zeta\zeta,
 \end{equation}
 and the equilibrium free energy, which is also the free energy at which the
entropy is a maximum, is given by the Legendre transform:
 \begin{equation}
 F = U - S + \mmu\cdot\overline\A + \mu_\zeta\zeta = 0.
 \end{equation}
 So we can either look for the maximum of the entropy, or equivalently we can
look for the zero of the free energy, to find the equilibrium values of $\A$
and $\zeta$.  In our particular calculations we maximized the entropy.

It is not necessary, as it is in some models, that the maximum entropy occur
at the point at which all the chemical potentials are zero.  The zero of the
chemical potentials in this model corresponds to the maximum of the entropy
{\em per layer}, which has no particular physical significance.  The
physically interesting quantity in this case is the entropy {\em per unit
volume}.  This quantity is not trivially related to the entropy per layer,
since different layers make different contributions to the volume of the
system.  The calculation actually performed involves finding the entropy per
layer and then converting it to an entropy density by dividing by the mean
volume per layer within the ensemble:
 \begin{equation}
 S_V = {S_{layer}\over a^2 \overline R_z},
 \label{soverv}
 \end{equation}
 where $\overline R_z$ is the $z$ component of the ensemble average of the
operator $\R(\sigma,\tau,l)$.  The mean volume per layer will vary with the
chemical potentials as we weight layers of different volumes more or less
heavily.  Because of the reflection symmetry of the ensemble in the $x$ and
$y$ axes, we do in fact expect the minimum entropy density to fall at $\mu_x
= \mu_y = 0$, but in general it will not also fall at $\mu_z = \mu_\zeta = 0$
and so we must probe a range of values of these chemical potentials to find
it, calculating the entropy density as shown above.





 \section{Results and discussion} Our calculation proceeds as follows.  We
define the transfer matrix $\T$ as in Equation~(\ref{defstm}) using the list
of possible transitions from one dead surface to another for a particular
system generated from the stacking graph of Paper~I.  The systems we have
looked at are the square-base canonical cell tilings with base of length
$\tau b$, $\tau^2 b$, and $\tau^3 b$ (known as the `$\frac21$', `$\frac32$',
and `$\frac53$' sizes).  Although the transfer matrix quickly becomes large
as the system size increases (the $\frac53$ system has a transfer matrix some
7000 elements on a side) it is very sparse, so that matrix multiplication can
be performed quickly, in time of order the rank of the matrix rather than the
number of elements.  The great advantage of the method we have adopted is
that the quantities we are interested can be calculated by evaluating only
the largest eigenvalue of the matrix, and its associated left and right
eigenvectors.  We can find these by simply multiplying the matrix many times
into a trial eigenvector.  Assuming this trial eigenvector has a non-zero
component in the direction of the lowest eigenstate of the matrix, this will
quickly give us the lowest eigenstate, and either one of the left or right
eigenvectors of the system.  We then repeat the process to find the other
eigenvector, and from these we can evaluate the free energy per layer of our
ensemble from Equation~(\ref{defsg}) and the free phason strain components
$\A$ from Equation~(\ref{htot}) as well as the parameter $\zeta$.  Using
these we can evaluate the entropy density per node from
Equation~(\ref{soverv}).

We then vary the values of the chemical potentials coupling to the phason
strain and $\zeta$ to find the maximum of the entropy for the ensemble, and
the quadratic variation about that maximum to extract the coefficients
$C_1$, $C_2$, and $C_3$, and hence the phason elasticities.  The most
straightforward way to accomplish this turns out to be to repeat the
calculation on a grid of values in the space of the three chemical
potentials that make up the vector $\mmu$, maximizing always with respect to
the remaining potential $\mu_\zeta$, as discussed in Section~III~D.  Then we
perform a cubic least-squares fit to the resulting function.  The quadratic
terms in this fit give us our elasticities, and the cubic terms give us an
estimate of the error involved in calculating second derivatives at the
maximum using what is essentially a finite difference method.

For the `$\frac21$' system, which has a base of length $\tau b$ and a
background phason strain of $\tau^{-5}$, we find that one of the horizontal
components of the phason strain---the one that we call $A_2$---can only take
one value no matter what sequence of layers we stack one on top of another.
Thus the corresponding elasticity parameter, $C_2$, is infinite.  The other
two parameters are finite.  For the `$\frac32$' system, which has a base of
length $\tau^2 b$ and a background phason strain of $\tau^{-7}$, all three
$C$ parameters have finite values, but the system is still sufficiently
constrained that the packing fraction of nodes, or equivalently the parameter
$\zeta$, takes only one value independent of the order in which we stack
surfaces to construct our tower, so that there is no possibility of allowing
this parameter to fluctuate and maximizing the entropy with respect to it.

\begin{figure}
 \begin{center}
 \psfig{figure=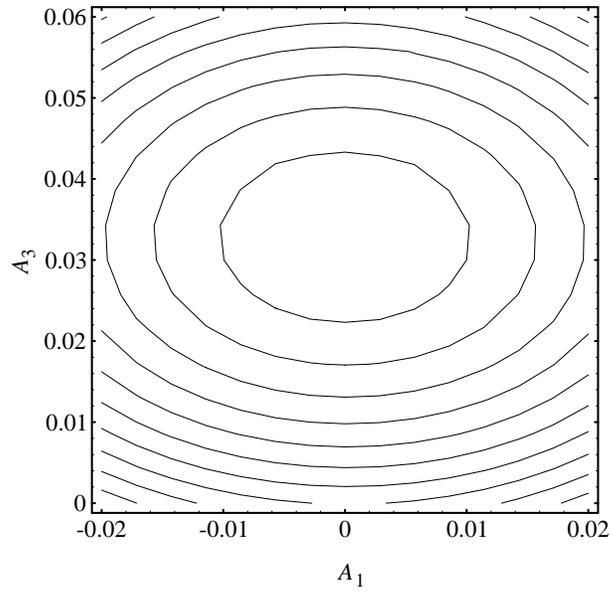,width=\figwidth}
 \end{center}
 \caption{Contour plot of the entropy as a function of the phason strain
components $A_1$ and $A_3$ in the region of the maximum entropy.}
 \label{entropy}
 \end{figure}

The first system large enough to exhibit the generic elastic behavior typical
of towers of canonical cells with large bases is the `$\frac53$' system,
which is also the largest system we have tackled.  In this system, which has
a square base of length $\tau^3 b$ on each side and a background phason
strain of $\tau^{-9}$, all three components of the strain vector $\A$ are
free to fluctuate under the influence of their corresponding chemical
potentials, as is the parameter $\zeta$ which measures the density of nodes
of the tiling.  Figure~\ref{entropy} shows a contour plot of the entropy
surface $\max_\zeta[S(\A,\zeta)]$ as a function of the phason strains $A_1$
and $A_3$.  The maximum entropy density is $6.00\times10^{-3}$, and falls at
$A_1=0$ and at a small non-zero value of $A_3$ ($=0.0077\ldots$), as
predicted.  (It also falls at $A_2=0$, though this is not evident from the
figure.)  We have also studied the dependence of the entropy on $\zeta$ and
we find that for a given $\E$, it depends very sharply on $\zeta$.  There is
a narrow valley in the plot of $f(\E;\zeta)$ centered around the plane
$\zeta=\zeta_0(\E)$.  In consequence, if we fix $\zeta$, the apparent
dependence of $f(\E)$ on $\E$ is very sharp.  (Note that, for the 2/1 and 3/2
cases, $\zeta$ is in fact a function of $\bf E$---the values of the
background phason strain are so high in those cases that the valley has zero
width.)

If we just omit (set to zero) the $\mu_\zeta$ parameter, thereby accepting
the ensemble which maximizes the entropy {\em per layer\/} at a given phason
strain, then we find an entropy density which differs from the actual value
by only about three parts in $10^5$.  We assume that the difference produced
by including terms in the Hamiltonian coupling to other densities of local
tiling patterns is smaller still, and therefore that we were justified in
neglecting them in this calculation.


The value of the parameter $\zeta$ at the maximum of the entropy for the
$\frac53$ ensemble was 0.334, which is close to the ``magic'' value of
$\zeta=0.317$ obtained at the end of Ref.~\onlinecite{henleyCCT}.  The
corresponding packing fraction $\phi$ for spheres of radius $c$ placed on the
nodes may be determined from formulas given in Sec.~III~A of
Ref.~\onlinecite{henleyCCT}.  After converting from the length units used
there, we obtain
 \begin{equation}
 \phi = { {\pi \sqrt 3}\over{16}} \biggl[{4\over{\sqrt 5}} \eta
            +2(1-\eta)(1-{\zeta\over3})\biggr].
 \end{equation}
 where
 \begin{equation}
 \eta \equiv {1\over 2}(1-\tau^9 \det\E)
 \end{equation}
 and $\det\E = B^2 A^{(0)}_3$ in our case.  For the $\frac53$
ensemble, we find a value of $\phi=0.606$.  Values for the other system
sizes are given in the table.

The packing fraction at which the maximum entropy occurs can be written
roughly as
 \begin{equation}
 \phi({\bf E}) = 0.6064 - 0.026 (A_3 - A_3^{(0)}) 
 \label{phizero}
 \end{equation}
 for the $\frac53$ case, with virtually no dependence on $A_1$ or $A_2$.  (It
seems plausible that, with more general values of phason strain, it would
depend on the determinant $\det\E$ of the strain tensor.)  In other words,
the optimal packing fraction is practically constant (varying $\sim1$\% over
the range of possible phason strains.)  In the $\frac21$ and $\frac32$ cases,
$\phi$ is a unique function of the phason strain.  In these systems the
coefficient of variation with $A_3$ is respectively 3.5 and 0.5 times as
large as in Eq.~(\ref{phizero}).


Values for the elastic parameters $C_1$, $C_2$, $C_3$, and for the phason
strain at equilibrium $A_3^{(0)}$ for the various systems are given in
Table~I.  The values do not fit the relations~(\ref{defsc}) very well,
presumably because the system size is small.  However, it is possible to make
out some trends in the data as the system size increases:

 \begin{itemize}
 \item[(i)] There is a clear tendency to oscillate between high and low values
of the elastic constants from one approximant to the next. 
 \item[(ii)] superimposed on this oscillation, there is also a strong tendency
for $C_m$ to decrease with increasing system size.
 \end{itemize}

The first of these trends is to be expected if the errors are caused by the
large background phason strain $B$, in view of Equation~(\ref{bps}).  The
same oscillation is seen in the Monte Carlo simulations of the rhombohedral
tiling by Shaw~{\it{}et~al.}~\cite{shaw1}.  (See, for instance, Figure~2 in
that paper.)

We suggest that~(ii) is also a finite-$B$ effect.  The phason strain tensor
components define a 9-dimensional space, and there is a domain in this space
(centered on zero) of allowed phason strains.  The entropy density vanishes
with singular derivatives at the boundary of this domain.  When $B$ is large,
the phason strain tensor is necessarily closer to the boundary of the allowed
domain, and the second derivatives of the entropy density (including the
coefficients $C_m$) are consequently larger.  In addition, the tendency for
$S_V$ to be larger for smaller systems also enhances $C_m$, because
increasing the maximum value of a function, while at the same time decreasing
the interval over which it must rise from 0 to its maximum and fall back to
zero, obviously forces it to have a larger (negative) second derivative.

So what can we say about the extrapolation of the data to infinite system
sizes?  The entropy density $S_V$ at least, appears to be far from its
asymptotic value in the systems we have studied here.  Indeed, on the basis
of these data alone, one could not really rule out $S_V=0$ in the
infinite-system limit~\cite{lowerbound}!


For the coefficients $C_m$, in view of the large values of $C_2$ for the 2/1
and $C_3$ for the 3/2, we have extrapolated the inverses $C_m^{-1}$.  We have
assumed the behavior \begin{equation} {1\over C_m^{(n)}} \approx {1\over
C_m^{(\infty)}} + \lambda B, \end{equation} where $C_m^{(n)}$ is the value of
$C_m$ measured for the $n^{\rm th}$ approximant and $B$ is background phason
strain.  For each $C_m$, we have extrapolated from the pair $(2/1,5/3)$ and
the pair $(3/2,5/3)$ and taken the mean of these as the extrapolated value,
writing error bars to span the two independent extrapolations~\cite{extrap}.
The results are given in the ``extrap'' column of the Table.


We have also performed a slightly different analysis of the $5/3$ system 
using just the 316 approximants described in Paper~I.  This calculation is
detailed in Appendix~A, and in the column marked ``$\frac53$ (F) in
the table.

Each of these different estimates give us four quantities $C_1$, $C_2$,
$C_3$, and $A_3^{(0)}$, but there are only two independent parameters $K_1$,
$K_2$ to fit to them.  This places constraints on the four quantities with
which their measured values are not entirely consistent.  However, we can
certainly conclude that $K_1 \approx 1$, and that $K_2$ is negative and
fairly large (possibly close to the instability limit~(\ref{stability})).
This sign of $K_2$ is opposite to that in the rhombohedral tiling, a fact
that we can explain qualitatively by the following argument.

From~(\ref{defsc}) we have
 \begin{equation}
 K_2 = {C_1-C_2\over\tau^2-\tau^{-2}}
 \end{equation}
 so that $K_2<0$ if and only if $h_x$ fluctuates more than $h_y$ as one walks
along a path in the $z$ direction.  Examining the tiling, we find that the
linkages along such a path are dominated by the 2-fold $b$ linkage that runs
entirely along $z$, and by the 3-fold $c$ linkage $(\pm \tau,0,\tau^3)$.  In
our basis~(\ref{perp}), these $c$ linkages have perp space components only in
the $x$ directions, making the $x$-fluctuations large compared with the $y$
ones, which in turn makes $K_2$ negative.
On the other hand, the linkage vectors in the rhombohedral tiling are just
the basis vectors in~(\ref{parallel}); thus the dominant steps in the
direction of the path are $\eta (0, \pm 1, \tau)$ which has fluctuations in
the $h_y$ direction rather than the $h_x$ direction.  Thus for that tiling
$K_2>0$~\cite{tang90,shaw1}.

For applications of random tilings to structural fitting, the parameter of
greatest interest is the effective phason Debye-Waller factor $\B\pe$, which
corresponds to the extra variance in perp space due to the randomness of the
tiling.  The experimental value of $\B\pe$ is of the order of 1\AA$^2$.  In
Appendix~B, using the ensemble of 316 packings~\cite{newman1}, we find a
crude theoretical estimate of $\B\pe$ for the canonical-cell random tiling
which is somewhat smaller.

Finally, we would like to consider whether there is the possibility of
performing a calculation for a larger system.  Probably there is.  It would
tax the power of the available computing resources, but it should be
possible.  We estimate (see Paper~I) that the next largest size of
system---the $\frac85$ system---should have a stacking graph of about one
million dead surfaces.  Given that the average number of nodes added between
one dead surface and another tends to a constant as the system becomes large,
we expect the time taken to construct the graph to scale as the number of
dead surfaces, and so we would require about 200 times as much CPU time to
perform the calculation as we did for the $\frac53$ system.  With today's
high-performance computing resources, such a calculation would be just within
the bounds of possibility.

 \section{Conclusions} Using the results of a previously-presented method
for breaking down and cataloging three-dimensional random tilings, we have
defined a transfer matrix whose eigenstates describe the properties of an
ensemble of random-tiling configurations in the shape of towers with a
finite area base and infinite extent in the $z$ direction.  The ensemble has
a mean phason-strain and packing fraction controlled by chemical potentials
whose values we can choose.  The transfer matrix is sparse and so can be
efficiently multiplied into a trial eigenvector to find the largest
eigenvalue.  This gives us the free energy density and thus the entropy
density for the ensemble.  Maximizing this with respect to the components of
phason strain and the packing fraction we can find the equilibrium values of
these quantities, and the variation of the entropy density about the
maximum, which defines the phason elasticity constants which are related to
experimentally measurable quantities such as phason Debye-Waller factors,
and diffuse scattering close to Bragg peaks.

We have applied our method to the `canonical-cell
tiling'~\cite{henleyCCT} for towers with a square base with sides
taking the three smallest non-trivial lengths possible.  For the two
smaller of these systems, we find that the systems have additional
constraints making their elastic behavior different from the generic
behavior expected of large systems.  The largest system we have
studied, which we refer to as the `$\frac53$' system is the smallest
one that exhibits the same elastic behavior as a large system.  From
this system we have extracted a value of $6.00\times10^{-3}$ for the
entropy per unit volume.  The two fundamental elastic constants, $K_1$
and $K_2$, are over-determined by the transfer matrix calculation
because elasticities corresponding to the strains in various
directions are related by symmetries of the system.  This gives us a
way to estimate the (finite-size) errors in our calculation, and it
turns out that the values for $K_1$ and $K_2$ are not very accurate
for this smallest of systems.  Our best estimate is that $K_1$ lies in
the range 0.8 to 1.1, and $K_2<0$.  In addition, the requirement that
the tiling be elastically stable with respect to other phases means
that $K_2/K_1>-\frac35$.


It is interesting to compare our results for the canonical-cell tiling with
those for the rhombohedral random tiling~\cite{tang90,shaw1}.  In doing so,
we must be careful to allow for the different lengths of the linkages making
up these tilings.

In the rhombohedral tiling~\cite{strandburg91} the entropy per node $S_N$ is
$\sim 0.24\pm0.02$, whereas in the canonical-cell tiling~\cite{cfhenley} it
is only $\sim0.047\pm0.01$.  This is a result of the fact that the
canonical-cell tiling is more constrained in the states it can take than the
rhombohedral tiling.  This constraint is also responsible for limiting the
canonical-cell tiling to only 32 distinct local node
environments~\cite{henleyCCT}.  By contrast, the node environments of the
rhombohedral random tiling are chosen from $10\,527$ of different
possibilities~\cite{baake}.

To compare the elastic constants of the two tilings, we first define a
characteristic unit of phason strain $E_0$ to be the ratio of the perp- and
real-space displacements of the fundamental linkage.  $E_0$ is 1 and
$\tau^{-3}$ for the rhombohedral and canonical-cell tilings, respectively.
Then a dimensionless measure of the elastic constant $K_1$ is the
corresponding decrease in free energy per node for a phason strain of
magnitude $E_0$:
 \begin{equation}
 \Delta F_N = \frac12 K_1 E_0^2/n,
 \end{equation}
 where $n$ is the packing fraction of nodes.  For the rhombohedral tiling,
$n = 0.6498$, $K_1 = 0.81\pm 0.01$ and $K_2 = 0.495\pm0.02$, hence $\Delta F_N
= 0.263$; for the canonical-cell tiling, using the extrapolated values $\Delta
F_N = 0.28$, very similar to the rhombohedral case~\cite{rhombs}.

Although the results for the elastic properties of the small canonical-cell
systems show signs of being strongly influenced by finite-size effects, we
believe that it should be feasible using modern supercomputing resources
to perform a similar calculation for a larger system which should yield a
more accurate estimate of the elastic constants for the infinite random
canonical-cell tiling.  We also believe that the method we have presented,
which is fundamentally different from transfer matrix methods previously used
in the study of random tilings, should be applicable to any random-tiling
model in any number of dimensions.  Even if the canonical-cell model does not
ultimately turn out to be as good a model of real icosahedral phases as some
other (yet to be proposed) tiling, we believe this method will prove useful
in the calculation of experimentally measurable properties of quasicrystals.

 \section{Acknowledgments} The authors would like to thank Mark Oxborrow and
Marc de Boissieu for useful discussions.  This work was supported in part by
the DOE under grant number DE--FG02--89ER--45405.

 \appendix
 \section{Fourier mode analysis} In the past, elastic constants have been
extracted from Monte Carlo simulations of finite random-tilings with periodic
boundary conditions, through measurement of the equilibrium fluctuations of
the phason strain.  We have carried out such an analysis for the present
system, to provide independent (albeit inferior) estimates of the elastic
constants.  In lieu of a simulation, we have used the {\it exact\/}
expectations for the ensemble with 316 microstates computed in Paper~I, which
are packings of a cube with side $L=\tau^3 b$.

The elastic free energy~(\ref{freal}) can be expanded around a state of zero
phason strain and rewritten as a sum over Fourier modes,
 \begin{equation}
 F_{tot} = \sum _\q \h(-\q) \K(\q) \h(\q)
 \end{equation}
 where $\h(\q)$ is the Fourier transform of the height field~(\ref{coarse})
normalized as in Ref.~\onlinecite{shaw1}.
 In the basis of~(\ref{parallel}) and~(\ref{perp}), the stiffness
coefficients are~\cite{jaricnelson}
 \begin{equation}
 K_{ij}(\q) = [K_1 |\q|^2 - K_2 [({1\over 3}|\q|^2 +\tau^{-1} q_{i+1}^2 -\tau
        q_{i+2}^2)] \delta_{ij} - 2q_i q_j (1- \delta _{ij} ).
 \end{equation}
 It then follows that
 \begin{equation}
 \langle h_\alpha(\q) h_\beta(\q) \rangle = \{ \K ^{-1} \}_{\alpha\beta}.
 \end{equation}

In order for $\h(\r)$ to be well defined in a cell with periodic boundary
conditions, we must adopt rationally-related perp space vectors as used
in~\cite{oxborrowthesis}.  In the present case, the perp space basis vector
$\eta(\tau,1,0)$ is replaced by $\eta'(5,3,0)$ where $\eta' \equiv (5^2 +
3^2)^{-1/2}$, and similarly for the other vectors in Equation~(\ref{perp}).

For our Fourier transform we used the crude definition
 \begin{equation}
 \h(\q) = {{V^{1/2}}\over N} \sum _i \h_i {\rm e}^{{\rm i}\q\cdot \r_i}
 \end{equation}
 following Ref.~\onlinecite{shaw1}.  (Interpolations of $\h(\r)$ between tile
vertices were used by Refs.~\onlinecite{oxborrowthesis}
and~\onlinecite{jeongsteinhardt}.)


We report results only for the smallest wavevector $\q=(0,0,2\pi/L)$, since
even this is too large to truly be the long-wavelength limit.  For that $\q$
value, $\K$ only has diagonal elements and they are simply $K_{ii}=|\q|^2
C_i$ with $C_i$ given by Equation~(\ref{defsc})~\cite{surprising}.  Finally
we obtain the estimates $C_i = |\q|^{-2} \langle|\h(\q)|^2\rangle^{-1}$ which
were reported in the table.  (The results from the next larger $\q$ value,
$(2\pi/L, 2\pi/L, 0)$, are also consistent with $K_2<0$, and $K_2/K_1$ being
of order unity, but with $K_1$ and $K_2$ each increased by a factor of at
least 2.)

 \section{The phason Debye-Waller factor} To define the perp-space
Debye-Waller factor, one assumes that there exists an ideal, perfectly
quasiperiodic structure made of canonical cells and that our random tiling
(when represented as a surface in 6-dimensional space) differs from this ideal
structure by random displacements in the perp-space direction, which have a
Gaussian distribution with each component having variance $2\B\pe$.  It
follows that each actual structure factor is reduced from its ideal value by
the factor $\exp(-\B\pe |{\bf G^\perp}|^2)$, where ${\bf G^\perp}$ is the
perp-space component of the reciprocal lattice vector.

Fitting of experimental data found $\B\pe=0.39$\AA$^2$ for
i(AlCuLi)~\cite{jaricqiu} and $\B\pe=0.70$\AA$^2$ for
i(AlPdMn)~\cite{boissieu1} (in units where the basis vectors in
Eqs.~(\ref{defsr}) and~(\ref{defsh}) have length $a_R=5.1$\AA\ and $4.65$\AA\
respectively.)  In the dimensionless units used in this paper, $\B\pe\approx
0.015$ and $0.034$, respectively.  If the variance of the microscopic
coordinates~(\ref{defsh}) of the nodes in the ideal structure is $\langle
|{\bf r^\perp}|^2 \rangle_0$, and in the actual structure their variance is
$\langle |{\bf r^\perp}|^2 \rangle$, then
 \begin{equation}
 \langle |{\bf r^\perp}|^2 \rangle-\langle |{\bf r^\perp}|^2 \rangle_0 =
 6\B\pe 
 \label{perpvar}
 \end{equation}
 exactly.

Using Eq.~(\ref{perpvar}), we have estimated $\B\pe$ from the ensemble of 316
packings of the cube of side $\tau^3 b$ (the so-called `$\frac53$'
approximants) constructed in Paper~I.  Since the quasiperiodic ideal
structure is unknown, $\langle |{\bf r^\perp}|^2 \rangle_0$ is unknown.
However, the packing in our ensemble with minimum perp-space variance
$\langle |{\bf r^\perp}|^2 \rangle_{min}$ has a high symmetry and is probably
a true approximant of the quasiperiodic structure; the perp-space variance of
this structure should not be too different from that of the ideal infinite
structure.  We find $\langle |{\bf r^\perp}|^2 \rangle_{min} = 0.2154$ and
$\langle |{\bf r^\perp}|^2 \rangle = 0.2758$.  Hence
 \begin{equation}
 \B\pe \approx {1\over 6} \langle |{\bf r^\perp}|^2 \rangle-\langle |{\bf
r^\perp}|^2 \rangle_{min} \approx 0.010
 \label{DWtheory}
 \end{equation}
 This is significantly smaller than the experimental value reported for
i(AlCuLi), and much smaller than that for i(AlPdMn).

\begin{table}
{\footnotesize
\begin{tabular}{|l|c|c|c|c|c|c|c|c|}
\hline
size & edge & $B$ & $C_1$ & $C_2$ & $C_3$ & $A_3^{(0)}$ & $\phi$ & $S_V$\\
\hline
\hline
$2/1$ (TM)   & $\tau b$   &  0.090 & 0.890 & $\infty$ & 1.282 & 0.0185 & 0.6004 & 0.0292 \\
$3/2$ (TM)   & $\tau^2 b$ & -0.034 & 2.101 & 5.835    & 10.14 & 0.0590 & 0.6051 & 0.0103 \\
$5/3$ (TM)   & $\tau^3 b$ &  0.013 & 0.452 & 2.302    & 0.494 & 0.0077 & 0.6064 & 0.0060 \\
extrap.      & $\infty$   &  0     & $0.48\pm0.08$ & $1.54\pm0.35$ &
$0.48\pm0.10$ &
               \vtop to 0.6cm{\hbox{0.014\hfil}\vfill\hbox{\hfil$\pm0.008$}} &
               \vtop to 0.6cm{\hbox{0.607\hfil}\vfill\hbox{\hfil$\pm0.001$}} &
               \vtop to 0.6cm{\hbox{$0.0041$\hfil}\vfill\hbox{\hfil$\pm0.0007$}} \\
$5/3$ (F)   & $\tau^3 b$ &  0.013 & 0.205 & 2.663    & 0.645 & 0.013  & 0.606 & 0.0036 \\
formula      & $\tau^{n-2}b$ & $\tau^{-(2n+3)}$ & $K_1+1.285 K_2$ &
               $K_1-0.951 K_2$ & $K_1-0.333 K_2$ & 
               ${-2B K_2\over K_1-0.333 K_2}$ & -- & -- \\
\hline
\end{tabular}
}
 \caption{Results for the elasticity coefficients and equilibrium phason
strain of various sizes of system.  The rows labeled TM were calculated from
the transfer matrix method, and the row marked MC was calculated by analyzing
the 316 cubic $\frac53$ approximants from Paper~I using the method employed
by Shaw~{\it{}et~al.}~\protect\cite{shaw1} in their Monte Carlo studies of
the square-triangle tiling (see Appendix~A).  The ``extrap.'' row is an
extrapolation of the transfer matrix results to a system of infinite size.}
 \label{results}
 \end{table}

\begin{references}
\bibitem{choice}
 This choice guarantees that, when expressed in
Cartesian coordinates, the phason elastic tensor has a cyclic symmetry under
$(xyz)$ permutations.
\bibitem{tisone}
 Notice that we use reduced (dimensionless) units of energy throughout,
without showing temperature explicitly.
\bibitem{chempots}
 However, so long as we have only a finite number of chemical potentials, we
do not have the exactly correct ensemble.  This caveat applies to all random
tiling calculations where volume per layer is variable.
\bibitem{isingtm}
 Our dead surface representation is reminiscent of a transfer-matrix approach
to Ising models on lattices with spiral boundary conditions in which just one
spin is added at a time~\cite{nightingale}, and has similar advantages.
\bibitem{muphi}
 In fact we implemented this part of the calculation as a chemical potential
coupling to the packing fraction $\phi$, rather than to $\zeta$ directly,
since the packing fraction is easier to evaluate.  Since $\zeta$ and $\phi$
are linearly related (see Ref.~\onlinecite{henleyCCT}), the two approaches are
equivalent.
\bibitem{extrap}
 There are various ways of extrapolating,
of course.  The most successful are:
 \begin{itemize}
 \item[(i)] fitting the ratio $C_1/C_2$: this obeys quite closely the form
$x+yB$, yielding $C_1/C_2=0.235\pm 0.010$.  (In recent experiments,
de~Boissieu~\cite{boissieu2} and co-workers have measured the anisotropy of
the diffuse scattering surrounding the Bragg peaks for i(AlPdMn) and fitted
their results to a model such as our having two independent elastic
constants.  They find values for these constants equivalent to $C_1/C_2 =
0.194$, which is in fair agreement with this figure.)  The value of $C_1/C_2$
is sufficient to determine a value $K_1/K_2$.  Unfortunately, $C_3/C_2$ is
too ill behaved to be useful.
 \item[(ii)] extrapolating $S_V/C_m$; the factor of $S_V$ helps cancel
out the trend denoted (ii) in the text.
 \end{itemize}
 Within the errors on our final result, the different extrapolations are
all in agreement.
\bibitem{cfhenley}
 This is only 40\% of the value previously quoted by one of the authors in
Refs.~\onlinecite{henleyART} and~\onlinecite{henley89}.  That earlier value
was just the $3/2$ entry in the Table.
\bibitem{rhombs}
 An alternative way of comparing the two tilings is to apply the standard
subdivision of the canonical-cell tiling into rhombohedra~\cite{henleyCCT}.
The entropy density of this sub-ensemble is $S_V + 3n_b \ln 2$ where $n_b$ is
the density of $b$ bonds, since each $b$ bond allows two local choices for
subdivision into rhombohedra~\cite{henleyCCT}.  This comes out to only half
of the total entropy of the unrestricted rhombohedron
tiling~\cite{strandburg91}; on the other hand, the elastic constant $K_1$ is
nearly the same for the two tilings.  The ``binary tiling''~\cite{binary}
also has nearly the same elastic constant as the unrestricted rhombus tiling,
of which it is a subensemble~\cite{widometal89}.
\bibitem{surprising}
 This is not so surprising, since the $C_i$ govern the fluctuations of
$\h(\r)$ along the length of the tower in the $z$ direction, which are all
sums of Fourier modes with $\q$ parallel to $\hat z$.
\end{references}
 \end{document}